\documentstyle[12pt]{article}

\begin{document}

\title{The Afterglow of GRBs} 
\author{Arnon Dar\\
Department of Physics \\
Technion-Israel Institute of Technology \\
 Haifa 32000, Israel}
\maketitle

\centerline{{\bf Abstract}}
Models where relativistic jets from merger or accretion induced
collapse in compact binary systems produce cosmological gamma ray bursts
(GRBs) also predict that GRBs are accompanied by delayed emission of high
energy photons, TeV neutrinos, X-rays, optical photons and radio waves.
Their emission mechanisms are similar to those responsible for their
emissions by blazars during gamma ray flares. The recently observed
afterglow of GRB 970228 in X-rays and optical photons is well accounted by
such models. In particular, the predicted power-law spectrum, time delay
which is inversely proportional to frequency and a power-law decay with
time which is energy independent, are all in excellent agreement with the
observations. 

\section*{Introduction} 
The origin of gamma ray bursts (GRBs), which were discovered more than 25
years ago, has been a complete mystery (e.g., Fishman and Meegan 1995).
Their observed isotropy in the sky, deficiency of faint bursts and the
lack of concentrations towards the Galactic center, in the Galactic disk
and in the direction of M31 strongly suggest (e.g., Briggs 1995) that they
are cosmological in origin (Prilutski and Usov 1975; Usov and Chibisov
1975; van den Bergh 1983; Paczynski 1986, Goodman 1986; Goodman, Dar and
Nussinov, 1987; Eichler et al. 1989) but no conclusive evidence for that
was available. However, this have been changing dramaticly since February 
28.
\noindent 
On Feb. 28.458 UT, eight hours after the gamma ray detectors
aboard the Compton Gamma Ray Observatory, Ulysses space craft, Wind
spacecraft and BeppoSAX satellite have detected the gamma ray burst GRB
970228 within a narrow error box (Costa et al. 1977a; Palmer et al.  1997;
Hurley et al. 1997), the BeppoSAX satellite detected in the $0.5-10~keV$
energy range an X-ray counterpart, i.e., a fading X-ray source at the same
position which was not known before (Costa et al. 1997b).  The field was
observed again on March 3.734 by BeppoSAX and on March 7.028 -7.486 by
ASCA in the $2-10~keV$ range (Yoshida et al. 1997) and the source was
detected at the same position at a flux level lower by a factor $\sim$ 20
and $\sim 35$, respectively.  An optical fading remnant was also detected
21 hours after the detection of the GRB (Groot et al. 1997) with
magnitudes $m_V=21.3$ and $m_I=20.6$ in the V and I bands, respectively. 
The fading optical GRB remnant was acquired by the Hubble Space Telescope
(HST) on March 26 which resolved it into a point-like source at 25.7
magnitude (Shau et al. 1997a) and an extended object (a faint galaxy ?).
It was detected again by HST on April 7 with $m_V=26.0\pm 0.3$ and
$m_I=24.6\pm 0.3 $ (Shau et al. 1997b). In the R band, where the 
relative contribution of the faint galaxy is expected to be larger,
the remnant was  detected with a slower decline, i.e.,
with  $m_R\sim 24$  between March 6.32 -13.0, by Keck II, INT,
Palomar 5m and ESO NTT, and  with $m_R\approx 24.9\pm 0.3 $ 
around April 6 by Keck II (e.g., Metzger et al. 1997). 
\bigskip
\noindent 
The identification of the site of GRB 970228 with a very faint galaxy
places it at a cosmological distance (although no absorption features that
could allow redshift determination have been reported). Cosmological
distances ($D\sim 10^{28}cm$) imply that GRBs have enormous energies,
$E\sim10^{51-52}\Delta\Omega~erg$, during short periods of time, where
$\Delta\Omega$ is the solid angle into which their emission is beamed. 
This enormous energy-release over a short time suggests that merger or
accretion induced collapse (AIC) of compact objects (neutron stars and
black holes) in close binary systems produce cosmological GRBs (Blinnikov
et al. 1984; Goodman et al. 1987; Eichler et al. 1989). Relativistic
beaming is required then in order to avoid self absorption through
$\gamma\gamma\rightarrow e^+e^-$. Relativistic fireballs (Cavallo and Rees
1978, Paczynski 1986; Goodman 1986) or relativistic jets (e.g., Shaviv and
Dar 1995a, 19996) therefore were invoked in order to explain GRBs. 
Observations and theoretical considerations favor relativistic jets as
their drivers.  Observations seem to indicate that all systems which are
powered by mass accretion onto a compact object release a large fraction
of their accretion power in kinetic energy of highly relativistic jets.
These relativistic jets are observed to produce very efficiently photons
whose energies extend from radio waves all the way up to TeV
$\gamma$-rays. The required energy release is more moderate in the jet
geometry. Therefore, Shaviv and Dar (1995, 1996) developed a model for
cosmological GRBs where highly relativistic hadronic jets in dense stellar
regions upscatter stellar light into gamma rays. 

\noindent 
Hadronic collisions of the relativistic jet particles with gas targets
(stellar wind, bloated stellar atmospheres, molecular cloud) can produce
high energy gamma rays, high energy neutrinos, and high energy electrons
and positrons. It was suggested that these high energy electrons and
positrons produce, through synchrotron emission and inverse Compton
scattering, afterglows in X-rays and optical photons of $\gamma$-ray
flares from blazars (Dar and Laor 1997) and of GRBs (Shaviv and Dar, 1996;
Dar 1997). In this letter we show that emissions from a jet can account
very well for the observed afterglow of GRB 970228 in X-rays and optical
photons. 

\section*{GRBs From Accretion Jets} 
The formation, of highly relativistic jets are not well understood.
However, the final accretion rates and the magnetic fields which are
involved in mergers/AIC of compact stellar objects are much larger than
those encountered in active galactic nuclei (AGN), microquasars and high
mass compact X-ray binaries. Therefore, it is very likely that
highly relativistic jets are also produced in mergers/AIC of compact
stellar objects. In view of the uncertainties in modeling mass ejection
driven by merger/AIC, rather Shaviv
and Dar (1996) assumed that NS/AIC produce jets with typical Lorentz
factors, $\Gamma\sim 10^3$, which are required by GRBs, beaming angles
$\Delta\Omega>\pi/\Gamma^2$ similar to those observed/estimated for AGN
and microquasars, and ejected mass $\Delta M\sim (dM/d\Omega)\Delta
\Omega\leq 10^{-3} M_\odot$ (because the released kinetic energy is
bounded by $E_K=\Gamma \Delta Mc^2 < M_\odot c^2 $). The natural birth
places of close binary systems are dense stellar regions (DSR) such as
galactic cores, star-burst regions and super star-clusters. These DSRs are
rich in gas and in very luminous stars which emit very strong stellar
winds (e.g., Tamura 1996). They have rather large photon and gas column
densities, $N_\gamma$ and $N_p$, respectively, larger than
$10^{23}cm^{-2}$. Resonance upscattering of interstellar photons in such
DSR can explain the fluence, duration complex light curves and spectral
evolution of GRBs, provided that the particles in the accretion jets are
injected with approximately a simple power-law distribution of Lorentz
factors, $dn/d\Gamma\sim \Gamma^{-\alpha}$ with $2\leq\alpha\leq 3$,
above a minimum value $\Gamma_m\sim 300$ (Shaviv 1996; Shaviv and Dar
1996). 

\noindent  
In fact, for a free ballistic expansion the ejected jet is optically 
thick to interstellar photons if $\Delta M>m_A R^2\Delta\Omega/
\sigma\approx 10^{-4}M_\odot\Delta\Omega$, where $R\sim 5\times 10^{17}cm$
is the typical size of DSR and $\sigma\sim 10^{-18}cm^2$ is the typical
resonance scattering cross section of optical photons from typical
mergers/AIC hadronic jets. Therefore, the ejected jet acts as a highly
efficient relativistic mirror which upscatters stellar photons along its
trajectory into beamed $\gamma$-rays. The typical photon fluence and
energy of GRBs produced by merger/AIC in DSR are 
 \begin{equation}
F_\gamma \approx  N_\gamma R^2/D^2\sim 10^{1\pm 1}~cm^{-2},          
 \end{equation} 
 \begin{equation} 
E_\gamma \approx \sigma N_\gamma(\Gamma\epsilon/ m_{A}c^2)  
E_K \sim  10^{51\pm 1}\Delta\Omega~erg.        
 \end{equation} 
  
\section*{High Energy Gamma Rays} 

The observed emissions of GeV $\gamma$-rays (e.g, Thompson et al. 1995)
and TeV $\gamma$-rays (e.g., Kerrick et al. 1995; Gaidos et al. 1996) from
blazars are usually interpreted as produced by inverse Compton scattering
of highly relativistic electrons in their jets on soft photons, internal
or external to the jet (Dermer and Schlickeiser 1993, 1994;  Blandford and
Levinson 1994). However, the kinetic energy of relativistic jets that are
made of normal (hadronic) matter resides mainly in protons and nuclei.
Therefore, the above mechanisms seem unable to convert enough kinetic
energy of hadronic jets into high energy photons and explain the observed
large fluxes of GeV and TeV $\gamma$-rays. Hadronic jets, however, can
convert a significant fraction of their kinetic energy into GeV-TeV
$\gamma$-rays through $pp\rightarrow \pi^0X$; $\pi^0\rightarrow 2\gamma$
in collisions with gas targets of high column density along the line of
sight, which are present near accreting massive black holes (Dar and Laor
1997). Gas targets with high column densities (stellar winds, bloated
atmospheres) can also be present near accreting stellar compact objects or
in star-burst regions and galactic cores where NS merge/AIC are expected
to be highly enhanced. They may convert a significant fraction of the jet
kinetic energy into high energy $\gamma$-rays. 

\noindent
The cross section for inclusive production of high energy $\gamma$-rays
with a small transverse momentum, $cp_{T}=E_{T}<1~ GeV$ in pp collisions
(e.g., Neuhoffer et al. 1971; Ferbel and Molzon
1984) is well represented by
 \begin{equation} 
{E\over \sigma_{in}} {d^3\sigma\over d^2p_{T}
dE_\gamma}\approx (1/2\pi E_0^2)e^{-E_{T}/E_0}~f_\gamma(x), 
 \end{equation} 
where $E$ is the incident proton energy, $\sigma_{in} \approx 35~mb$ is 
the $pp$ total inelastic cross section at TeV energies, $E_0\approx 0.16~
GeV$ and $f_\gamma (x)\sim (1-x)^3/\sqrt{x}$ is a function only of the 
Feynman variable $x=E_\gamma/E$, and not of the separate values of the 
energies of the incident proton and the produced $\gamma$-ray. The 
exponential dependence on $E_{T}$ beams the $\gamma$-ray production into 
$\theta <E_{T}/E\sim 1/6\Gamma$ along the incident proton direction. 
When integrated over transverse momentum the scaled inclusive cross section 
becomes $\sigma_{in}^{-1}d\sigma/ dx\approx f_\gamma (x).$ If the protons in 
the jet have a power-law energy spectrum, $dF_p/dE\approx AE^{-\alpha}$, 
then, because of Feynman scaling, the produced $\gamma$-rays have the 
same power-law spectrum:  
 \begin{equation} 
{dF_\gamma\over dE}
\approx N_p \sigma_{in} \int_{E}^{\infty} {dF_p\over dE_p}{d\sigma
\over dE}dE_p 
\approx N_p\sigma_{in}I_\gamma AE^{-\alpha}, 
 \end{equation} 
where $N_p$ is the column density of the target and $I_\gamma
=\int_0^1x^{\alpha-1}f_\gamma(x)dx $.  Gas targets with large column
densities, $N_p>10^{23}cm^{-2}$, are present in/near galactic cores and in
star burst regions. For instance, such regions
seem to be very enriched in variable luminous blue supergiants (e.g.,
Tamura 1996). Such stars whose luminosity exceeds $\sim 10^5L_\odot$,
also eject mass at a rate of up to $L_m\sim 10^{-4}M_\odot~yr^{-1}$ with
an outflow speed of $v\sim 500~km~s^{-1}$. If such a star is located at a
distance $R_*\sim 5\times 10^{17}~cm$ from the explosion and at an angle
$\theta_*\leq 1/\Gamma$ relative to the jet, its wind produces an
effective gas target inside the beaming cone with an average column
density of $N_p\approx L_m/2m_pv R_*\theta_*\sim 10^{23}~cm^{-2}$, which
can convert $\sim 10^{50\pm 1}erg$ of the jet energy into high energy
$\gamma$-rays. 

\section*{High Energy Neutrinos} 
Hadronic production of photons in diffuse targets is also accompanied by
neutrino emission through $pp\rightarrow\pi^{\pm}\rightarrow
\mu^{\pm}\nu_\mu$ ; $\mu^{\pm} \rightarrow e^{\pm}\nu_\mu\nu_e $.  For a
proton power-law spectrum, $dF_p/dE= AE^{-\alpha}$ with a power index of
$\alpha\sim 2$, one finds (e.g., Dar and Shaviv 1996) that the 
produced spectra of $\gamma$-rays and $\nu_\mu$'s satisfy 
\begin{equation} 
dF_{\nu_\mu}/dE \approx N_p\sigma_{in}I_{\nu_\mu} AE^{-\alpha} 
\approx 0.7 dF_\gamma/dE. 
\end{equation} 
Consequently, we predict that high energy $\gamma$-ray emission from GRBs
is accompanied by emission of high energy neutrinos with similar fluxes,
light curves and energy spectra. At energies above TeV, $\gamma$ rays from
distant GRBs are strongly attenuated by the infrared cosmic background
radiation (e.g., Stecker et al. 1993).  However, the universe is
transparent to TeV neutrinos. Therefore, detection of TeV neutrinos can be
used to confirm the hadronic nature of jets and to detect distant GRBs in
TeV emission.  The number of $\nu_\mu$ events from a GRB in an
underwater/ice $\nu_\mu$ telescope is $SN_AT\int
R_\mu(d\sigma_{\nu\mu}/dE_\mu)(dF_\nu/ dE)dE_\mu dE$, where $S$ is the
surface area of the telescope, $N_A$ is Avogadro's number, $T$ is the
duration of the GRB, $\sigma_{\nu_\mu}$ is the inclusive cross section for
$\nu_\mu p \rightarrow \mu X$, and $R_\mu$ is the range (in $ gm~cm^{-2}$)
of muons with energy $E_\mu$ in water/ice. For a GRB with $F_\gamma \sim
10^{-6}~erg~cm^{-2}$ above $E_\gamma=1~TeV$ and a power index $\alpha=2$,
we predict 3 neutrino events in a $1~km^2$ telescope. If the observed GeV
gamma ray emission from GRB 940217 (Hurley et al. 1994) extended into the
TeV region it could have produced hundreds of $\nu_\mu$ events in a
$1~km^2$ neutrino telescope. 
 
\section*{X-Rays, Optical Photons and Radio Waves}
The production chain $pp\rightarrow
\pi^{\pm}\rightarrow\mu^{\pm}\rightarrow e^{\pm}$ that follows jet-gas
collisions enriches the jet with ultrarelativistic electrons. Due to
Feynman scaling, their differential spectrum is proportional to the
emitted high energy $\gamma$-ray spectrum, 
 \begin{equation} 
dF_e/dE \approx N_p\sigma_{in}I_e AE^{-\alpha}
\approx 0.4 dF_\gamma/dE.
 \end{equation}
If they encounter a perpendicular magnetic field they cool radiatively by
synchrotron emission and inverse Compton scattering and their initial
energy decreases according to $\Gamma=\Gamma_0/(1+t/t_{1/2})$ with
$t_{1/2}(\Gamma_0)=3mc/4\sigma_T\Gamma_0(u_{B}+u_\gamma)$, where $\sigma_T
$ is the Thomson cross section and $u_{B}= B_\perp^2/8\pi$ and $u_\gamma$
are the energy densities of the transverse magnetic field and the
radiation field, respectively.  The acceleration, the collimation and the
little deflection of jet particles by interstellar magnetic fields,
as well as the emission of polarized radiations by relativistic jets,
indicate that strong magnetic fields are present within jets.  Similarly,
strong radiation fields may also be be present within jets due to the
beamed emission of synchrotron radiation (inverse Compton scattering from
synchrotron photons of energy $\epsilon$ can upscatter them to energy
$E_\gamma\sim (4/3)\Gamma_e^2\epsilon$ in the observer frame),
Consequently, for $\alpha\geq 2$, radiative cooling by inverse Compton
scattering and synchrotron emission changes the initial power-law energy
spectrum of the ultrarelativistic electrons into
 \begin{equation}
dF_e/dE \approx N_p\sigma_{in}I_e AE^{-\alpha}(1-t/t_{1/2}(E))^{\alpha-2}. 
 \end{equation} 
Thus, for $\alpha>2$ the electron spectrum cuts off at
$\Gamma_e\approx 3m_ec^2/4(u_{B}+u_\gamma)\sigma_{T}ct$ 
and the photon emission cuts off at a corresponding frequency.  
For $t\ll t_{1/2}$ the electron spectrum maintains its original
spectral form $\sim E^{-\alpha}$. Synchrotron emission and  
inverse Compton scattering from electrons
with such a power-law spectrum produce a power-law 
spectrum:  
 \begin{equation}
 dF_\gamma/dE\sim E^{-(\alpha+1)/2}.  
 \end{equation}   
The radiative cooling time of jets must be very long (jets from
microquasars reach hundreds light years, while AGN jets reach
hundred thousands light years before disruption). However,
the magnetic (and the radiation) energy density 
within the ejected jets that expand ballisticly decreases like
$u_{B}=B^2/8 \pi\sim R^{-3}\sim t^{-3}$ where $R$ is the distance
from the ejection point and $t$ is the time in the observer frame.
Maximum emission of synchrotron radiation by electrons with energy
$E_e$ occurs at photon energy 
$E_\gamma\sim B_\perp E_e^2.$   
Photons with energy $E_\gamma$ at time $t$ are produced
mainly by electrons with energy $E_e\sim(E_\gamma/B_\perp)^{1/2}\sim 
E_\gamma t^{3/2}$. Because the cooling time of jet particles is 
relatively long, the electron spectrum is 
proportional to $ E_e^{-\alpha}$ and consequently  
the decay of the afterglow is independent of frequency 
and is given approximately by 
 \begin{equation}
d^2F_\gamma /dEdt\sim E^{-(\alpha+1)/2}t^{-3(\alpha-1)/4}.    
  \end{equation}   
Eq. (9) also follows from the fireball model of Wijers et al. 1997. 
It is valid until inverse Compton scattering of the
microwave background photons, which produces the same power-law 
spectrum but decays very slowly with time, takes over.
The model also predicts a time lag $t_d$ in the appearance of the
afterglow in different energy bands, which is inversely proportional to
photon energy. This time lag is due to the difference in times of flight
of electrons with different energies along the jet, as seen by the
observer, which is proportional to $1/\Gamma_e^2$, i.e., $t_d\sim 
1/E_\gamma$. 
\noindent 
The measured afterglow in X-rays and optical photons of GRB 970228 at
different observations times during the first 38 days are very well
described by eq.(9) with $\alpha\approx 2.6$, i.e., by $Ed^2F_\gamma
/dEdt\sim E^{-0.8}t^{-1.2}$ (see, e.g., Wijers et al. 1997). 

\section*{Conclusions} 
If GRBs are produced by merger/AIC jets in star burst regions or galactic
cores of distant galaxies, they are also accompanied by delayed emission
(afterglow) of high energy photons, TeV neutrinos, X-rays, optical photons
and radio waves. Their predicted spectral and temporal behavior are
correlated and follow simple power-laws, as described by eq.(4) for high
energy gamma rays, by eq.(5) for TeV neutrinos, and by eq.(9) for X-rays,
optical photons and radio waves, respectively. The predictions are in
excellent agreement with the observed afterglow of GRB 970228 in X-rays
and optical photons. 
  
\bigskip
\centerline{{\bf References}}

\noindent 
Blandford, R.D. \& Levinson, A. 1995, ApJ,  441, 79 

\noindent 
Blinnikov, S.I. et al. 1984, Sov. Ast. Let. 10, 177 

\noindent 
Briggs, M.S. Ap. \& Sp. Sc. 1995, 231, 3 

\noindent 
Cavallo, G. and Rees, M.J. 1978, MNRAS, 183, 359

\noindent 
Costa, E. et al. 1997a, IAU Circ. 6572 

\noindent 
Costa, E. et al. 1997b, IAU Circ. 6576 

\noindent 
Dar, A., 1997, in ``Very High Energy Phenomena In The Universe'',
Proc. XXXII Rencontre De Moriond, Les Arcs, France,
January 19-24, 1997.  

\noindent 
Dar, A. \& and Laor, A., 1997, ApJ, 478, L5    

\noindent 
Dar, A. \& Shaviv, N. 1996, Astropar. Phys.. 4, 343 

\noindent 
Dermer, C.D. \& Sclickeiser, R. 1993, ApJ, 415, 418

\noindent 
Dermer, C.D. \& Sclickeiser, R. 1994, ApJS, 90, 945

\noindent 
Eichler, D. et al. 1989, Nature, 340, 126 

\noindent 
Ferbel, T. \& Molzon, W.R. 1984, Rev. Mod. Phys. 56, 181

\noindent 
Fishman, G.J. and  Meegan, C.A.A. 1995, Ann. Rev. Ast. Ap. 33, 415 

\noindent 
Gaidos, J.A. et al. 1996, Nature, 383, 319

\noindent 
Goodman, J. 1986, ApJ, 308, L47 

\noindent 
Goodman, J., Dar, A. and Nussinov, S. 1987,  ApJ, 314, L7 

\noindent 
Groot, P.J. et al. 1997, IAU Circ. 6584

\noindent 
Hurley, K., et al. 1994, Nature, 372, 652

\noindent 
Hurley, K., et al. 1997, IAU Circ. 6578

\noindent 
Kerrick, A.D. et al. 1995, ApJ, 438, L59   

\noindent 
Meszaros, P. and Rees, M.J. 1992, MNRAS, 257, 29 

\noindent 
Metzger, M.R., et al. 1997, IAU Circ. 6631  

\noindent 
Neuhoffer, G., et al. 1971, Phys. Let. 37B, 438   

\noindent 
Paczynski, B. 1986, ApJ, 308, L43 

\noindent 
Palmer, D., et al. 1997, IAU Circ. 6577

\noindent 
Prilutski, O.F. and Usov, V.V. 1975, Ap\&SS, 34, 395

\noindent 
Shau, K., et al. 1997a, IAU Circ. 6606

\noindent 
Shau, K., et al. 1997b, IAU Circ. 6619

\noindent 
Shaviv, N.J. 1996, Ph.D Thesis  (Technion Report Ph-96-16).

\noindent 
Shaviv, N.J. and Dar, A. 1995, ApJ, 447, 863 

\noindent 
Shaviv, N.J. and Dar, A. 1995, MNRAS, 277, 287 

\noindent 
Shaviv, N.J. and Dar, A. 1996, Submitted for publication in PRL;
See also Proc. of VI Rencontre De Blois, Blois, France June 5-11, 1996,
in press.  

\noindent 
Stecker, F.W., et al. 1993, ApJ, 415, L71

\noindent 
Tamura, M. et al. 1996,  ApJ, 467, 645   

\noindent 
Thompson, D.J., et al. 1995, ApJS, 101, 259 

\noindent 
Usov, V.V. and Chibisov, G.B. 1975, Sov. Ast. 19, 115
 
\noindent 
van den Bergh, S. 1983, Ast. \& Ap. Suppl. 97, 385 

\noindent 
Wijers, R.A.M.J., et al. 1997, submitted to MNRAS (astro-ph 9704153)   

\noindent 
Yoshida, A., et al. 1997 IAU Circ. 6593 

\end{document}